\documentclass{article}
\usepackage{amsmath, amsfonts}
\usepackage{graphicx}
\usepackage{natbib}

\title{Multivariate distance matrix regression for a manifold-valued response variable}

\author{
	Matthew Ryan\\
	\and
	Gary Glonek\\
	\and
	Melissa Humphries\\
	\and
	Jono Tuke\\
}
\renewcommand{\vec}[1]{\mathbf{#1}}
\DeclareMathOperator{\chol}{chol}
\newcommand{\RR}{\mathbb{R}}
\newcommand{\tr}{\mathrm{tr}}
\newcommand{\keywords}[1]{\textbf{Keywords:} \textit{#1}}

\begin{document}
	\maketitle
%
%
	\begin{abstract}
		In this paper, we propose the use of geodesic distances in conjunction with multivariate distance matrix regression, called geometric-MDMR, as a powerful first step analysis method for manifold-valued data.  Manifold-valued data is appearing more frequently in the literature from analyses of earthquake to analysing brain patterns.  Accounting for the structure of this data increases the complexity of your analysis, but allows for much more interpretable results in terms of the data.    To test geometric-MDMR, we develop a method to simulate functional connectivity matrices for fMRI data to perform a simulation study, which shows that our method outperforms the current standards in fMRI analysis.
	\end{abstract}

	
	\section{Introduction}\label{sec:Intro}
	

	The process of finding a relationship between a variable of interest $y$ and a set of possible explanatory variables $x_1, x_2, \ldots, x_p$ is a fundamental notion in statistics.  When exploring this relationship, accounting for any structure one may find inherently in the data allows for more accurate and directly interpretable results.  For instance, if one of our explanatory variables $x_i$ is an ordinal categorical variable this should be accounted for.  This is equally true when there is structure found in our response variable $y$, such as interesting geometrical properties when $y$ is manifold valued, that is, $y$ can naturally be viewed as a point on a Riemannian manifold.
	
\begin{figure}
		\centering
		\includegraphics[width=\linewidth]{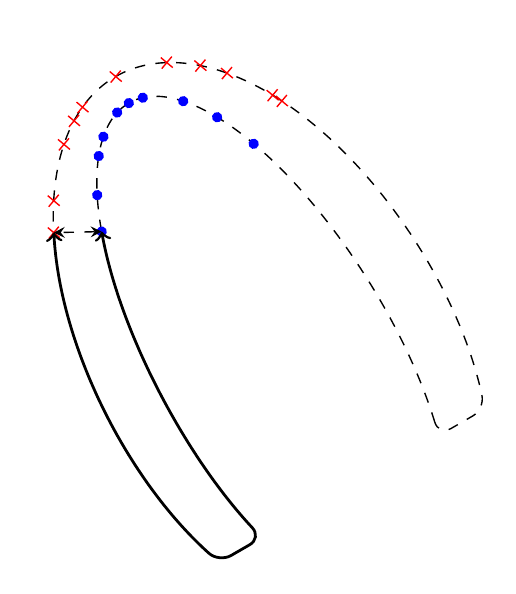}
		\caption[horseshoe diagram]{An example where considering the natural geometry of the data is the right thing to do. Here we have two sets of data points, the circles and the crosses. If considered as points in $\RR^2$, you would find it difficult for distances between the points to distinguish the groups (the short, dashed line). However, when considered as points on the horseshoe shape this distinction in distance becomes clear (the long, bold line).}
		\label{fig:horseshoe-diagram}
	\end{figure}

	The reason you would account for this geometric structure is not always immediately obvious in higher dimensions, but the idea can be highlighted in 2-dimensions.  Consider the situation in Figure \ref{fig:horseshoe-diagram}.  Here you see 2-dimensional response data that belongs to a natural horseshoe-like shape.  If you consider the Euclidean geometry between these points, it is difficult to detect any differences between the two groups as they are interspersed and clustered.  However, if you account for the geometry by travelling along the horseshoe, you see a clear differentiation between the groups as they suddenly become very far apart.
	
	Accounting for the geometrical structure in response variables has appeared in a wide variety of  applications, from studying patterns of earthquakes \cite{Cohen2015} to analysing trends in neuroimaging data \cite{Pennec2006,  Venkatesh2020}.  This is usually done by either constructing statistical methods on a Riemannian manifold to be applied in a general framework, such as geodesic regression \cite{Fletcher2013}, or exploiting the Riemannian structure inherently found in the data of interest, such as Venkatesh \textit{et. al.}'s work on participant identification \cite{Venkatesh2020}.	A key reason to generalise these Euclidean methods to Riemannian manifolds is that it allows us to model complex non-linear relationships in the data in a more interpretable manner \cite{Fletcher2013}.  This notion of modelling complex non-linear relationships in an interpretable manner ties in nicely with the intention of multivariate distance matrix regression (MDMR) \cite{Anderson2001, Mcardle2001, Zapala2006}.
	
	MDMR, otherwise known as PERMANOVA, is a subject-oriented analysis method which aims to associate observed differences in the response variables of subjects (as defined by a pairwise dissimilarity matrix $D$) to a given set of predictors.    First developed by Anderson and McArdle \cite{Anderson2001, Mcardle2001} for use in ecological data, MDMR has been used in areas ranging from bioinformatics \cite{Zapala2006} to neuroimaging \cite{Ponsoda2017, Shehzad2014}.  The theory of MDMR has been well developed since its inception \cite{ Anderson2013,  Anderson2003, Zapala2012}, and it has proven itself as an effective method for determining associations in data with a large set of response variables such as gene expression data \cite{Zapala2006} or functional magnetic resonance imaging (fMRI) \cite{Ponsoda2017, Shehzad2014}.   This makes MDMR an attractive alternative to multivariate ANOVA because the results are valid when there are more response variables than there are subjects in the study, since the object under consideration is the dissimilarity matrix.

	In this paper we propose a novel approach to MDMR for manifold-valued response data, which we call geometric-MDMR.	  Geometric-MDMR accounts  for the geometry of the data through the use of a geodesic distance for the dissimilarity matrix $D$.  When you have manifold-valued data, you can consider how far apart these data are as elements on the manifold by considering paths of minimal local length (geodesics) between the points.   Since the geodesics on the manifold are precisely determined by the chosen geometry, this will respect the geometrical properties of the data.  We show that our method has more power to detect group differences than simply using Euclidean distances through a simulation study.  We also argue that Geometric-MDMR is intuitively more interpretable in terms of the data.

	\section{Method}\label{sec:Methods}
	
	Throughout this paper, the following notation is used:
	
	\begin{itemize}
		\item $N$ denotes the number of subjects under consideration.  The letters $q$ and $p$ will denote the number of response and predictor variables respectively.
		\item Bold, lower case letters such as $\boldsymbol{x}$ and $\boldsymbol{y}$ denote vectors.
		\item $d$ will denote a distance or dissimilarity measure between the response variables.
		\item Upper case letters such as $A$ and $G$ denote matrices.
		\item A superscript $T$ such as $A^T$ will denote the transpose of a matrix.
		\item A matrix $A$ can be defined by its $ij^{th}$ element with the notation $A = \left[ a_{ij}\right]_{ij}$.
	\end{itemize}
	
	\subsection{MDMR}\label{sec:MDMR}
	
Multivariate distance matrix regression (MDMR)  \cite{Anderson2001, Mcardle2001,  Zapala2006} is an alternative approach to multivariate ANOVA for testing hypotheses on high-dimensional data.   MDMR provides a permutation-based test for ANOVA-like hypotheses through the calculation of a pseudo-F statistic.  The key difference between MDMR and multivariate ANOVA is that MDMR focuses on the pairwise dissimilarity matrix of the response variables and the relationship this has with the predictors.

Consider data  $(\vec{x}_1,\vec{y}_1), (\vec{x}_2 , \vec{y}_2), \dots, (\vec{x}_N, \vec{y}_N)$, where $\vec{y}_i \in \RR^q$ is a q-variate response variable, and $\vec{x}_i  = (1, x_{1i}, x_{2i}, \dots, x_{pi})$ is a vector of $p$-predictors and an intercept.  Let $d$ denote a dissimilarity or distance measure on the $\vec{y}_i$ and let $D = \left[d(\vec{y}_i, \vec{y}_j)\right]_{ij}$ be the pairwise dissimilarity matrix.   Denote by $X$ the matrix whose $i^{th}$ row is $\vec{x}_i$.   Consider the double-centred Gower matrix \cite{Gower1966} $G$ given by 
\[
		G = \left(I - \frac{1}{N} \vec{1} \vec{1}^T\right) A \left(I - \frac{1}{N} \vec{1} \vec{1}^T\right)\, , 
\]
where $I$ is the $N\times N$ identity matrix, $\vec{1} \in \RR^N$ is a vector of ones, and $A = \left[ -\frac{1}{2} d(\vec{y}_i, \vec{y}_j)^2\right]_{ij}$.  Then the pseudo-F statistic is given by
\[
	\tilde{F} = \dfrac{\tr(H G H)/(N-p-1)}{\tr( (I-H)G(I-H))/(N-1)}\, ,
\]
where $H = X(X^TX)^{-1}X^T$ is the usual projection matrix.  The motivation behind the pseudo-F statistic is that when $d$ is the Euclidean distance and $q = 1$, then $\tilde{F}$ corresponds with the usual $F$ statistic from ANOVA.  Thus $\tilde{F}$ is a natural extension of the $F$ statistic to an arbitrary dimension and distance measure.


As mentioned in the introduction, one of the main strengths of MDMR over ANOVA is that it is valid for $q > N$ since it relies solely on the distance measure between the data.  This makes MDMR a valuable association test for analysing gene expression data or neuroimaging data, which are generally very high dimensional.

	\subsection{A geometric point of view} \label{sec:geodesics}
	
	Statistics and differential geometry are not new acquaintances, since manifolds appear very naturally in many fields of study \cite{Fletcher2013, Frechet1948, Pennec1999}.  Geometrically inspired analysis has found applications all over statistics, from detecting humans in photographs \cite{Tuzel2007}, to analyse the change of shape for parts of the brain \cite{Fletcher2013}, to modelling the spatial distribution of earthquakes \cite{Cohen2015}.  
	
	A key idea in the above works is analysing \textit{geodesics} on the respective geometries.   Geodesics are commonly described on manifolds and relate to the notions of distance and angles on the manifold, so they naturally depend on the geometry under study.
	
	A \textit{manifold} $M$ is a space that locally looks like Euclidean space and patches together in a nice way.  Generally speaking, a manifold is a space on which we can perform calculus.   We may consider tangent vectors on a manifold, which are encapsulated in the tangent bundle $TM$.  To measure angles between tangent vectors and distances between points on the manifold we need to choose a geometry by way of a \textit{Riemannian metric} $g$  on the tangent space.  A Riemannian metric is a smooth inner product on the tangent space, which makes sense of the ideas of angles and distances.
	
	Let $u, v \in M$ and $\gamma:[0,1]\rightarrow M$ be a path from $u$ to $v$.  Using the Riemannian metric, we can measure the length of $\gamma$ on $M$ as
	\[
		L(\gamma) = \int_0^1 \sqrt{ g(\dot{\gamma}(t), \dot{\gamma}(t)) }\, dt \, ,
	\]
	where $\dot{\gamma}$ is the derivative in time of $\gamma$. The path $\gamma$ is a geodesic if 
	\[
		L(\gamma) = \min_{\gamma' \in \Gamma}\int_0^1 \sqrt{ g(\dot{\gamma'}(t), \dot{\gamma'}(t)) }\, dt \,,
	\]
	where $\Gamma$ is the set of all paths (defined on $[0,1]$) from $u$ to $v$.  In fact, we define the \textit{Riemannian distance} between $u$ and $v$ as
	\[
		d(u, v)  = \min_{\gamma' \in \Gamma}\int_0^1\sqrt{g(\dot{\gamma'}(t), \dot{\gamma'}(t)) }\, dt \, .
	\]
	This notion of distance is the key idea to geometric-MDMR.  It allows us to perform MDMR on manifold valued data while accounting for this Riemannian geometry.

	\subsubsection{Example: $\mathbb{R}^n$}
	
	The simplest example of a manifold would be normal Euclidean space $\mathbb{R}^n$.  This space is trivially a manifold with tangent space  $\mathbb{R}^n$.  The canonical Riemannian metric on this space is given by the dot product between vectors, which induces the usual Euclidean geometry and distance $d(\vec{u}, \vec{v}) = \| \vec{u} - \vec{v}\|_2$.  The geodesics that achieve this distance are straight lines through $\mathbb{R}^n$.

	\subsubsection{Example: $S^2$}
	
	The 2-sphere $S^2 = \{ (x, y, z) \in \mathbb{R}^3 : x^2 + y^2 + z^2 = 1 \}$ is a $2$-dimensional manifold, which can be shown through the use of stereographic projection.  The tangent space for a point $\vec{v} \in S^2$ is the plane that touches $S^2$ at exactly the point $\vec{v}$.  We can define the standard Riemannian metric on $S^2$ as the restriction of the Riemannian metric from $\mathbb{R}^3$ to $S^2$.  Under this metric, the geodesics on $S^2$ are given by the arcs on the great circles, and the distance between two points is the length of these arcs.

	\subsubsection{Example: $S_n^+$}
	
	Consider the space of positive definite symmetric $n \times n$ matrices
	\[
		S_n^+ = \{ A \in \mathrm{GL_n\mathbb{R}} : A = A^T, \, \boldsymbol{\alpha}^T A \boldsymbol{\alpha}> 0 \text{ for all } \boldsymbol{\alpha} \in \mathbb{R}^n\backslash\{\boldsymbol{0} \} \}\, ,
	\]	
	which is the natural home of covariance and correlation matrices.  $S_n^+$ is an $\dfrac{n(n +1)}{2}$-dimensional manifold with tangent space at $A\in S_n^+$ given by $T_A S_n^+ = \{ B \in \mathrm{GL_n\mathbb{R}} : B = B^T\}$.   To define a Riemannian metric on this manifold is non-trivial.  Following \cite{Forstner2003, Pennec2006}, we consider the affine-invariant geometry.  Let $B_1, B_2 \in T_{A}S_n^+$, then the affine-invariant Riemannian metric is given by
	\[
		g_A(B_1, B_2) = \mathrm{tr}\left( A^{-\frac{1}{2}} B_1 A^{-1} B_2 A^{-\frac{1}{2}} \right)\, .
	\]
	The geodesic distance between $B_1, B_2 \in S_n^+$ is given by:
	\begin{equation}
		d(B_1, B_2) = \| \log(B_1^{-\frac{1}{2}}B_2 B_1^{-\frac{1}{2}}) \|_2 = \sqrt{\sum_{i = 1}^n \log(\sigma_i)^2}\, ,
		\label{eq:s+geo}
	\end{equation}
	where $\sigma_1, \sigma_2, \dots, \sigma_n$ are the eigenvalues of  $B_1^{-\frac{1}{2}}B_2 B_1^{-\frac{1}{2}}$.

	\section{Simulation Study}\label{sec:Sims}

Resting-state fMRI involves subjects lying inside an MRI scanner set to detect changes in the blood-oxygen-level-dependent (BOLD) contrast periodically over the scan time \cite{Ogawa1990}.  This results in hundreds of thousands of  volumetric pixels (voxels) representing spatial regions in the brain, with each voxel having a time series of possibly hundreds of observations.  Voxels are typically divided into regions of interest (ROI) which involves aggregating the information of many voxels into similar  regions across the brain, defined either functionally or anatomically. One then analyses the  functional connectivity matrix of the ROIs as defined by the Pearson correlation matrix between the respective time series. 

A typical resting-state fMRI study aims to detect differences in the functional connectivity of the brain between study groups, that is, they aim to detect if the brains of healthy control subjects function differently to patient groups.  To explore the effectiveness of our method in such a study, we create a simulation designed to replicate group differences in the functional connectivity between subjects.  In this sense, we can consider the response variable for each subject from a resting-state fMRI study as a correlation matrix (or functional connectivity matrix) that naturally lives on the space $S_R^+$, where $R$ is the number of regions in the ROI decomposition.
	

	
\begin{figure}
		\centering
		\includegraphics[width=\linewidth]{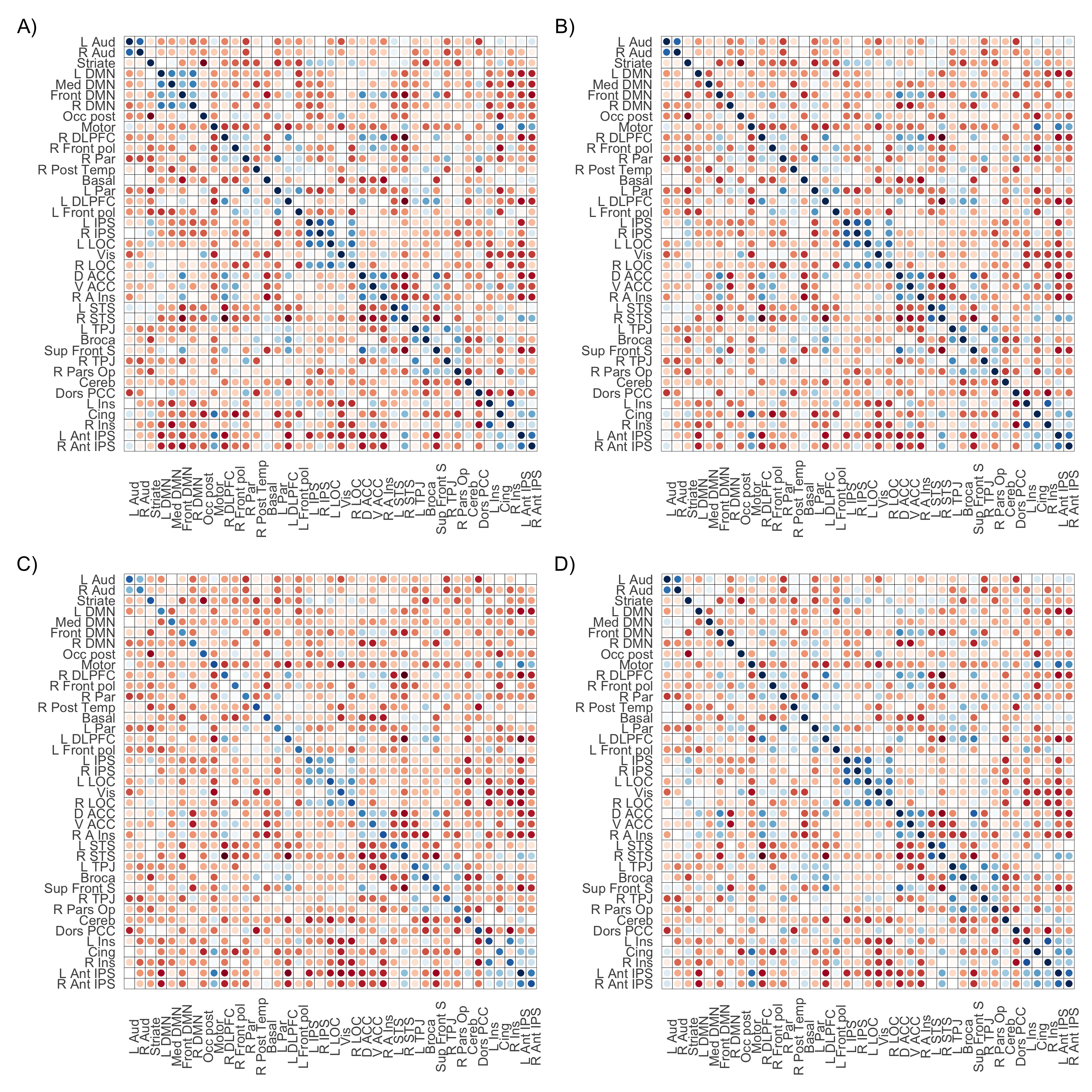}
		\caption[Explanation of simulation]{A visual representation of how to simulate the functional connectivity matrix for $b = 4$, $m = -0.55$, and $s = 0.267$. Figure A shows the randomly sampled subject from the real cohort of data (COBRE dataset). Figure B shows how the signal has been implanted into the DMN (upper left region). Figure C shows a simulation from the Wishart distribution using this matrix as the scale matrix. Figure D show the Wishart simulation normalised to a correlation matrix.}
		\label{fig:simulationexplanation}
	\end{figure}

	The process of simulating fMRI is non-trivial.  There are packages to simulate fMRI data in python \cite{Ellis2020}, MATLAB \cite{Erhardt2012}, and R \cite{Welvaert2011}; each of these packages focus on generating the functional time series for a given voxel in the brain.  Here, we are interested in the analysis of the functional connectivity matrix, and so propose a novel simulation method by constructing underlying functional connectivity matrices from real data and then perturbing them with known, implanted signals.  This is done by randomly sampling a functional connectivity matrix from a cohort of real subjects and implanting a signal into the functional connectivity matrix based on the simulated subjects group (either \lq\lq Patient\rq\rq\, or \lq\lq Control\rq\rq).  This matrix is then used as  the scale matrix in a Wishart distribution to simulate from the Wishart distribution.  We then normalise this simulation into a correlation matrix, and repeat this process for every subject.  This method is summarised in Figure \ref{fig:simulationexplanation}.
	
	For the real cohort of subjects, we chose to look at the COBRE dataset \cite{Aine2017}, which aims to explore differences in the functional connectivity between healthy controls ($n = 74$) and schizophrenic patients ($n = 72$).  The correlation matrices we consider are the $39 \times 39$ matrices defined by the MSDL atlas \cite{Varoquaux2011}.   This data comes with an array of phenotypic data to use as predictors, and for this simulation we focus solely on the subject group (\lq\lq Patient\rq\rq\, or \lq\lq Control\rq\rq).  More information on this dataset is found in Appendix A.

	If our simulated subject is a patient, we implant a signal of the form 
	\[
		B = \begin{bmatrix}
			1 & \rho & \rho^2 & \rho^3 & \cdots & \rho^{b-1}\\
			\rho & 1 & \rho & \rho^2 & \cdots & \rho^{b-2}\\
			\vdots & & \ddots &  & & \vdots\\
			\rho^{b-1} & \rho^{b-2} & \rho^{b-3} & \rho^{b-4}  &\cdots & 1
		\end{bmatrix}\, 
	\]
	into the default mode network (DMN).  The DMN is a collection of four particular ROIs in the MSDL atlas, and is a well studied functional network in the brain that is believed to be most active when a subject is awake and at rest \cite{Bijsterbosch2017}.  This functional network was chosen as it was found to have no significant association with the patient group in the COBRE dataset.  Note that the parameter $\rho$ controls the strength of the signal being implanted and the parameter $b$ controls the size of the signal, that is, how many consecutive ROIs we are implanting signal into.
	
	For these simulations, we consider $b = 2, 3$, or $4$, ranging from the smallest possible signal of interest ($b = 2$) to the size of the actual DMN in the MSDL atlas ($b = 4$).  We also consider $\rho =\tanh(p)$ where $p \sim N(m, s^2)$ for $m =-1.83, -0.55, 0, 0.55$, or $1.83$, and $s = 0.267$.  The values for $m$ were chosen to provide a  progression of correlations from $-0.95$ to $0.95$ on the $\mathrm{atanh}$ scale.  The value $s = 0.267$ was chosen as this is the standard deviation observed in the DMN for the COBRE dataset, on the $\mathrm{atanh}$ scale.

	The process of implanting this signal gives us a matrix $\tilde{A}(r)$ for $r \in [0,1]$.  This matrix is such that $\tilde{A}(0)$ is the original correlation matrix, $\tilde{A}(1)$ has the full signal $B$ implanted, and $\tilde{A}(r) \in S_R^+$ for all $r\in [0,1]$.  The specifics of the simulation method are explained in  Appendix B.

	We test the power of  geometric-MDMR against the current standards of MDMR used in neuroimaging \cite{Ponsoda2017, Shehzad2014}, which convert the functional connectivity matrices into vectors by taking the upper triangle, and use either Euclidean distance or a correlation-based distance on the derived vectors.  We use the group as the predictor in our MDMR.

\begin{figure}
		\centering
		\includegraphics[width=\linewidth]{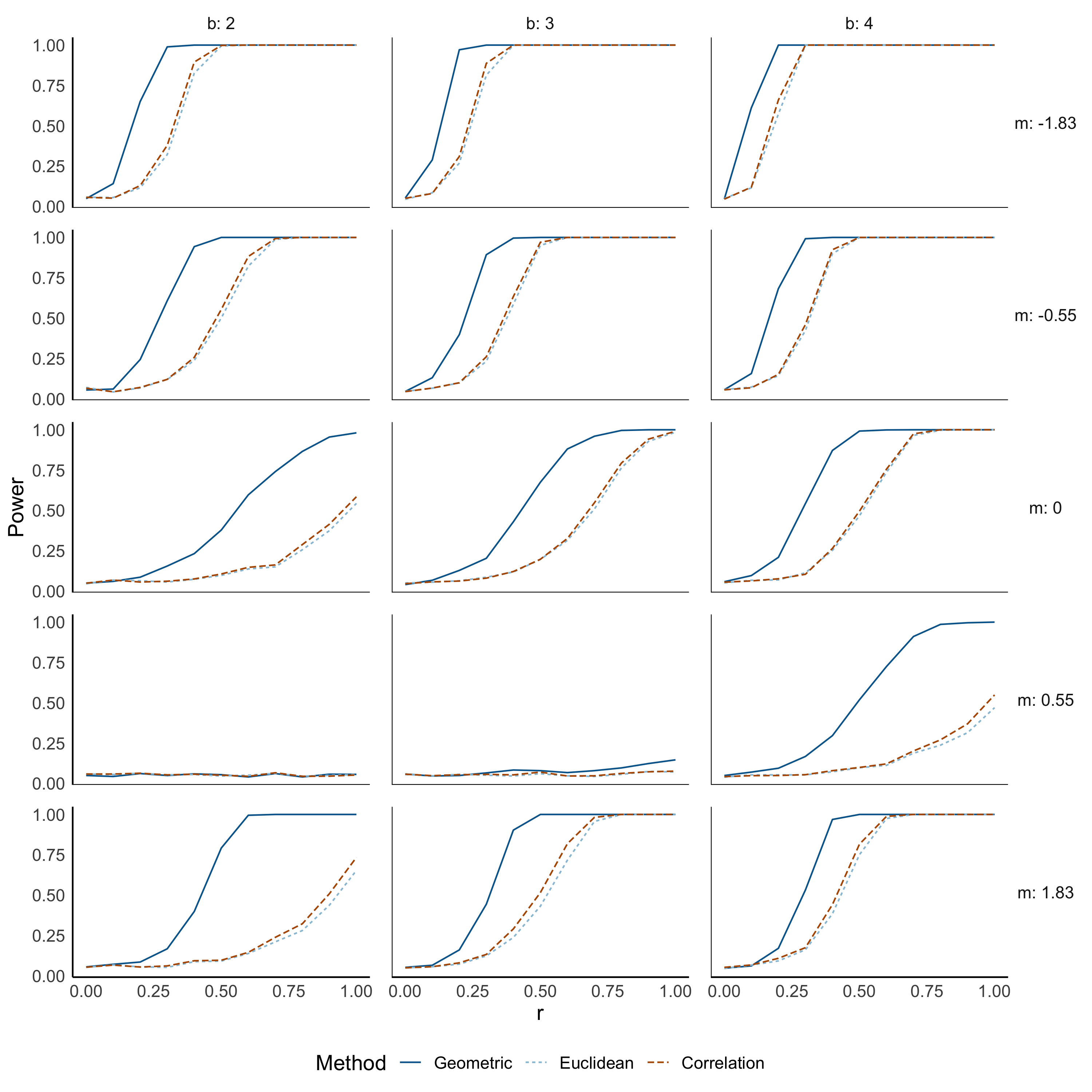}
		\caption{The results of the simulation power study comparing geometric-MDMR with the current standards in fMRI analysis.   For each combination of the parameters, we repeated the simulation 999 times and calculated the p-value from an MDMR for each of the three methods.  The $y$-axis represents the power from the simulation study as given by the proportion of simulations that returned a p-value of less than 0.05.  The $x$-axis represents the transition from no signal implanted ($r = 0$) to a full signal implanted ($r = 1$) for each set of parameters.  The columns represent different sizes of the signal as represented by $b$.  The rows are different strengths of the signal as represented by the mean of the normal distribution we simulate from $m$.}
		\label{fig:fcpowerplot}
	\end{figure}

	The results of the simulation study are seen in Figure  \ref{fig:fcpowerplot}, where you can clearly see that the geometric-MDMR outperforms the current standards.  By using the geometric extension, we find that the MDMR results become more sensitive to subtle changes in the data, likely because this method considers the data in its natural geometry rather than forcing a Euclidean structure.  In this example, the affine-invariant geometry pushes the matrices with zero or infinite determinants out to infinity, which is ideal as these would not be considered functional connectivity matrices.   This results in distances between functional connectivity matrices being stretched along a curved path, and we can think of the distance between them as being the distance between \textit{valid functional connectivity matrices}.
	
	\section{Discussion} \label{sec:Discussion}

	With the melding of a differential point of view (through the use of geodesics) and multivariate distance matrix regression, we have provided a powerful tool for an \textit{a priori} analysis on manifold-valued data.  Through simulations, we have shown that our method has increased power to detect significant differences in manifold-valued data over the current standards of embedding this data into Euclidean space.  We have also argued that geometric-MDMR is more interpretable  in terms of the data than the current standards.


	Geometric-MDMR would make a valuable contribution to any manifold driven analysis method as an \textit{a priori} test of association in the data.  This could allow researchers to easily reduce their predictor space by removing the predictors that are shown to have insignificant relationships as determined by geometric-MDMR, leading to simplified and stronger post-hoc analysis.

	It should be noted that multivariate distance matrix regression can only be used to determine if a relationship between the predictor variables and response variables exists, but not the nature of that relationship.  This makes geometric-MDMR an excellent first step for an analysis, but should be used in conjunction with other methods to strengthen the results.
	
	A drawback specific to considering geodesic distances is that it makes problems more mathematically complicated to formulate.  Take for instance the example of fMRI analysis.  The current standard when using MDMR \cite{Ponsoda2017, Shehzad2014} is to vectorise the upper triangle of the functional connectivity matrices, which is both simple to do and simple to conceptualise.  Comparing this to the formula in Equation \eqref{eq:s+geo}, it is clear how much more mathematically complicated geometric-MDMR can be.  However, we believe that this increase in difficulty is greatly outweighed by the stronger interpretability and power of geometric-MDMR.
	
	Geometric-MDMR provides an excellent first step to determine if a relationship is present in our data.  So far, post-hoc analysis of MDMR tends to be done on a situational basis, so post-hoc analysis for geometric-MDMR has yet to be explored.  A plausible candidate would be something like geodesic regression \cite{Fletcher2013}, so the conjunction of these two methods opens an avenue for further research.

	\bibliographystyle{authordate1}
	\bibliography{bib/library}
	\appendix
	
	\section{COBRE and the MSDL atlas}
	The data we consider in this paper is the COBRE dataset \cite{Aine2017}, which was downloaded using the \textit{Python} package \textit{nilearn v 0.6.2}.  This is an open source fMRI study that provides anatomical and functional MRI images for 72 patients with Schizophrenia and 74 healthy control patients.  The data was preprocessed using NIAK 0.17 under CentOS version 6.3 with Octave version 4.0.2 and the Minc toolkit version 0.3.18.  The data was also subjected to confound regression where they removed six motion parameters, the frame-wise displacement, five slow drift parameters, average parameters for white matter, lateral ventricles, and global signal, as well as 5 estimates for component based noise correction \cite{Behzadi2007}.  
	
	The ROI atlas we consider for this data is the multi-subject dictionary  learning (MSDL) atlas \cite{Varoquaux2011}.  This is a functional brain atlas, meaning that voxels are grouped together based on similar brain function instead of anatomical location.  This atlas partitions the brain into 39 functional nodes belonging to 17 distinct brain networks.  By brain network, we mean a collection of nodes that have been shown to work cohesively together.

	\section{Simulating functional connectivity matrices}
	This section will describe the simulation process from Section \ref{sec:Sims} in more detail.  Recall that to simulate a single subject we do the following:
	\begin{enumerate}
		\item Randomly select a correlation matrix $A$ from our chosen dataset (the COBRE data).
		\item If the subject is to be in the patient group, generate a signal matrix $B$ as described in Section \ref{sec:Sims}.  Implant this signal into $A$.
		\item Use the resulting matrix as the scale matrix for a Wishart distribution to produce a simulation for the subject.
		\item Normalise this matrix into a correlation matrix.
	\end{enumerate}
	Here, we describe how  the signal matrix $B$ is  implanted into the original matrix.  In the following, fix values for $b$ and $m$ from Section \ref{sec:Sims}.
	
	Suppose you have a correlation matrix $A$ (an $R\times R$ matrix) and wish to implant a signal $B$ (a $b \times b$ matrix).  Without loss of generality, suppose you are implanting $B$ into the top left corner of $A$ (you may rearrange the columns and rows of $A$ such that this is always true).  Write
	\[
		A = \begin{bmatrix}
				A_{11} & A_{12}\\
				A_{21} & A_{22}
			\end{bmatrix},
	\]
	where $A_{11}$ is of dimension $b \times b$.  Define 
	\[
		B(r) = (1-r) A_{11} + r B
	\]
	for $r \in [0,1]$, and 
	\[
		C(r) = \begin{bmatrix}
			\chol(B(r))^T ( \chol(A_{11})^{-1})^T & \vec{0}\\
			\vec{0} & I_{R - b}
		\end{bmatrix}
	\]
	where $\chol$ stands for the Cholesky square root, and $I_{R-b}$ is the $(R-b)\times(R-b)$ identity matrix.  Then
	\[
		\tilde{A}(r) = C (r)A C(r)^T
	\]
	has the property that $\tilde{A}(0) = A$, $\left[\tilde{A}(1)\right]_{11} = B$, and $\tilde{A}(r) \in S_R^+$ for all $r \in [0,1]$.  This is the process by which we implant the signal matrix $B$ into $A$.  If the simulated subject is not in the patient group, we take $\tilde{A}(r) = A$ for all $r \in [0,1]$.
	
	Now that you have a valid correlation matrix $\tilde{A}$ for a subject, you can add variational noise via the Wishart distribution.  That is, the simulation you consider for the subject is a random observation from a $\mathrm{Wishart}(\tilde{A}, k)$, where the degrees of freedom $k$ is randomly selected from integers between $39$ and $150$.  The lower bound $39$ is chosen as this is the dimension of $\tilde{A}$ in the COBRE dataset, and the upper bound of $150$ is chosen as this is the length of the time series observed for each subject in the COBRE dataset.  The resulting matrix is then normalised so that the diagonal entries are all one and it is a valid correlation matrix.
\end{document}